\documentclass[letterpaper,12pt]{article}
\usepackage[utf8]{inputenc}
\usepackage[inner=2.5cm,outer=2.5cm]{geometry}
\usepackage[english,activeacute]{babel}
\usepackage{amssymb,amsmath,alltt}
\usepackage{amsfonts}
\usepackage{enumitem}
\usepackage{graphicx}
\usepackage{authblk}
\usepackage{soul}
\usepackage{hyperref}

\usepackage[title]{appendix}
\usepackage{xcolor}

\definecolor{red}{RGB}{250,0,0}
\definecolor{purple2}{RGB}{120,0,120}
\definecolor{blue2}{RGB}{0,0,180}
\definecolor{green2}{RGB}{0,180,0}
\definecolor{grey2}{gray}{0.6}

\newcommand{\R}{\mathbb{R}}

\newcommand{\eps}{\varepsilon}

\newtheorem{thm}{Theorem}[section]

\title{Conefield approach to identifying regions without flux surfaces for magnetic fields}

\author[1]{D.Martínez-del-Río\footnote{david.martinezdelrio.math@gmail.com}}
\author[1]{R.S.MacKay}
\affil[1]{Mathematics Institute, University of Warwick, Coventry CV4 7AL, UK}

\date{}

\begin{document}

\maketitle

\begin{abstract}
{The conefield} variant of a Converse KAM method for 3D vector fields, identifying regions through which no invariant 2-tori pass transverse to a specified direction field, is tested on some helical perturbations of an axisymmetric magnetic field in toroidal geometry.
This implementation computes bounds on the slopes of invariant tori of a given class and allows to apply  a subsidiary criterion to extend the non-existence region, saving significant computation time.
The method finds regions corresponding to magnetic islands and chaos for the fieldline flow. 
\end{abstract}

%\keywords{
%}

\section{Introduction}
\label{sec:intr}
KAM theory establishes conditions under which invariant tori %\st{continue to}
exist in Hamiltonian systems. Specifically,  all but a small fraction of the invariant tori from generic integrable Hamiltonian systems remain when subjected to sufficiently small and smooth perturbations  \cite{dumas2014kam}.
Despite this, proving the existence of a significant portion of the tori seen in numerical simulations remains a challenging task, as discussed in \cite{figueras2017rigorous}.

An alternative method involves identifying regions where no invariant tori of a given class are present. This technique, referred to as \textit{Converse KAM theory} \cite{mackay1985converse,mackay1989criterion}, is much simpler to implement and gives close to optimal conclusions without excessive computation.

The present work presents a new application of Converse KAM theory to magnetic fields, complementing the results of \cite{KMM23}. First, it applies an alternative, yet equivalent, implementation of the method described in \cite{mackay2018finding}, called the ``conefield method''. This version of the Converse KAM method {constructs} bounds at any point for the slope of the tangent plane to any possible invariant torus through it that is transverse to a prescribed direction field.  Secondly, as discussed in the following sections, this approach enables processing the method's output data to further extend the region of non-existence by a subsidiary criterion called ``killends'', inspired by \cite{mackay1985converse}.

The paper illustrates the application of the method on the same  magnetic fields in toroidal configurations considered in \cite{KMM23}, with the same purpose of identifying regions where no invariant tori (flux surfaces) of a given class are present. A class of tori is defined by prescribing a direction field almost everywhere and requiring that the tori be transverse to this field. The primary choice for the direction field is the gradient of an appropriate squared distance function from a closed field line (magnetic axis),  with respect to a given metric.

In Section \ref{sec:magfields}, we introduce the magnetic fields tested in this paper. Section \ref{sec:cKAM} outlines the application of the Converse KAM method to  magnetic fields and presents the subsidiary criterion to enlarge the detected non-existence region. If preferred, Sections \ref{sec:magfields} and \ref{sec:cKAM} may be read in the opposite order.  In Section \ref{sec:results}, we present the results of the numerical implementation of the Converse KAM method and the subsidiary criterion on the chosen magnetic fields. Section \ref{sec:disc} presents a discussion of the results and section \ref{sec:conc} summarises our conclusions.

\section{Toroidal helical magnetic fields}
\label{sec:magfields}
In this paper, we choose to illustrate our alternative Converse KAM method on the same magnetic fields as  considered in \cite{KMM23}, which correspond to perturbations of a circular tokamak field by helical modes, based on \cite{kallinikos2014integrable}. For that reason, we give only a short exposition about the fields and the adapted coordinates used to represent them. A reader interested in further details is  referred to \cite{KMM23}.

Given a coordinate system $(x^1,x^2,x^3)$,  we will write a magnetic field $B$ in terms of its contravariant components $B^{i}$, rather than its physical ones. %; they differ by length factors 
%(see Appendix B in \cite{KMM23} for a summary about components of vector fields in curvilinear coordinates). 
This choice reduces the equations of motion for fieldline flow to simply $\dot{x}^i(t) = B^{i}(x(t))$; here ``time'' is to be understood to represent flow along the magnetic field lines at rate $|B|$.

The fields considered have 
a circular magnetic axis of radius $R_0>0$ in the horizontal plane $z=0$, which makes it easy to identify the ``principal'' class of tori that surround it.  They  are best described and treated in an adapted toroidal coordinate system $(\psi,\vartheta,\phi)$,  
which is a variant of the standard toroidal coordinates $(r,\theta,\phi)$. The latter coordinates are related to Cartesian coordinates $(x,y,z)$ through 
$$x = R \sin\phi,\ y = R\cos \phi,\ z = r\sin \theta,$$
where
$$R= R_0 + r\cos\theta,$$
for  $0\le r<R_0$.  $R$ represents the cylindrical radius relative to the $z$-axis. In these coordinates, the metric tensor is represented by the matrix $\text{diag}(1,r^2,R^2)$.

Following \cite{kallinikos2014integrable}, the adapted coordinates $(\psi,\vartheta)$  make the restriction $\beta_T$ of the magnetic flux-form $\beta$ to a poloidal section ($\phi=$ constant) take the form
\begin{equation}
\label{eq:tflux2}
\beta_{\text T}=d\psi \wedge d\vartheta,
\end{equation}
where $\wedge$ denotes the exterior product of differential forms (see \cite{mackay2020differential} for a tutorial),
 where $\psi$ is the toroidal magnetic flux across the poloidal disk of radius $r$ about a point of the magnetic axis, divided by $2\pi$. Thus, 
\begin{equation}
\label{eq:tflux1}
\beta_{\text T}=rR B^\phi\, dr \wedge d\theta.
\end{equation}
 For simplicity, we choose our magnetic fields to  have $B^\phi = {B_0R_0}/{R^2}$, which
corresponds to a vertical current along the $z$-axis.
Then $\vartheta$ can be constructed by equating (\ref{eq:tflux2}) and (\ref{eq:tflux1}), up to choice of origin that we take to be at $\theta=0$. 
Given the previous choices, we arrive at \cite{kallinikos2014integrable, abdullaev2014, KMM23}
\begin{align}
\begin{split}
&\psi= B_0R_0\left(R_0-\sqrt{R_0^2-r^2}\right)\\
&\tan \frac{\vartheta}{2} = \sqrt{\frac{R_0-r}{R_0+r}}\tan\frac{\theta}{2}.
\end{split}
\end{align}

\subsection{Magnetic fields studied}
To enforce volume-preservation by the fieldline flow, we specify $B$ as the curl of a vector potential $A$.  In terms of the covariant components of $A$, the contravariant components of $B$ are given by
$$B^{i} = \frac{1}{\sqrt{|g|}} \epsilon^{ijk} \partial_j A_k,$$
where $\epsilon$ is the Levi-Civita symbol and $|g|$ is the determinant of the matrix $g$ representing the metric tensor, $ds^2 = g_{ij} dx^i dx^j$. In our adapted toroidal coordinates, the volume factor $\sqrt{|g|}$ is $ 1/B^\phi=R^2/(B_0R_0).$  

We take a vector potential with helical modes introduced in its toroidal component, of the form {(in covariant components)}
\begin{align}
\label{eq:potential}
\begin{split}
A_\psi &= 0\\
A_\vartheta &= \psi \\
A_\phi &= -[w_1\psi+w_2\psi^2 + \sum_{m,n} \eps_{mn} \psi^{m/2} f_{mn}(\psi) \cos(m\vartheta-n\phi + \zeta_{mn})],
\end{split}
\end{align}
where $w_1\in\R$, $w_2\ne 0$, $m,n$ are integers with $m\ge 2$, $f_{mn}$ are smooth functions and $\zeta_{mn}$ arbitrary phases.  The factor $\psi^{m/2}$ is to make the resulting vector potential smooth at the coordinate singularity $\psi=0$.

The vector potential (\ref{eq:potential}) gives rise to the magnetic field $B=(B_0R_0/R^2) V$, where the (contravariant) components of the auxiliary vector field $V$ are
\begin{align}
\label{eq:fields}
\begin{split}
V^\psi &= \sum_{m,n} m\eps_{mn}\psi^{m/2} f_{mn}(\psi)\sin(m\vartheta-n\phi + \zeta_{mn})\\
V^\vartheta &= w_1 + 2 w_2 \psi + \sum_{m,n} \eps_{mn} \psi^{m/2-1} \left[\tfrac{m}{2}f_{mn}(\psi)+\psi f_{mn}'(\psi)\right]\cos(m\vartheta-n\phi + \zeta_{mn})\\
V^\phi &= 1 .
\end{split}
\end{align}
The cylindrical radius $R$ occurring in the conversion from $V$ to $B$ can be expressed in our adapted coordinates via
$$R = \frac{R_0^2-r^2}{R_0-r\cos\vartheta}$$
with $$r = \sqrt{2\frac{\psi}{B_0}-\frac{\psi^2}{B_0^2R_0^2}}.$$
%but we can avoid the conversion by applying the Converse KAM method to $V$ rather than $B$, as will be explained.

Because we take $m\ge 2$, the fields all have $\psi=0$ as a closed fieldline, as claimed, which we call the {\em magnetic axis}.

We define the {\em principal class} of tori to be the differentiable tori that are transverse to 
the vector field $\xi=\partial_\psi$.
%\st{$\nabla \psi$.}
%For example, with no helical modes the field is integrable with integral $\psi$  and the invariant tori $\psi=$ constant belong to the principal class.  So do all $C^1$-small deformations of them.
%\st{Specifying $\nabla \psi$ entails a choice of Riemannian metric, but there is no need to use the Euclidean one.  It is preferable to choose a metric so that $\nabla \psi$ is in the same direction as $\partial_\psi$.}  
Then the principal class of tori consists of the graphs of $\psi$ as a differentiable function of $(\vartheta,\phi)$.  %\textcolor{red}{[We don't actually choose a metric so perhaps this is an unnecessary diversion?]}
%\textcolor{purple2}{[Agreed. We still use $\xi = \partial_\psi$ but we do not do anything else, so we should shorten this paragraph.]}

Finally, the full magnetic flux-form $\beta$  is defined by $\beta = i_B \Omega$ where $\Omega$ is the volume form, or equivalently by $\beta = dA^\flat$, where $$A^\flat = A_\psi d\psi + A_\vartheta d\vartheta + A_\phi d\phi .$$   Thus 
$$\beta = V^\psi d\vartheta \wedge d\phi + V^\vartheta d\phi \wedge d\psi + V^\phi d\psi \wedge d\vartheta.$$
Because $V^\phi=1$, we see that restricted to a poloidal section, $\beta = \beta_{\text T} = d\psi \wedge d\vartheta$, as claimed earlier.

\section{Converse KAM for magnetic fields}
\label{sec:cKAM}
Let $B$ be a nowhere-zero $C^1$ vector field on an orientable $3D$ manifold 
$M$. In this paper, a
\emph{torus} in $M$ means a diffeomorphic image of the standard 2-torus $\mathbb{T}^2$, with $\mathbb{T}= \mathbb{R}/\mathbb{Z}$ (relaxation to Lipschitz tori will be discussed in section \ref{sec:disc}).  
We define a \emph{class} of tori in $M$  by specifying a $C^1$ vector field $\xi$, called ``direction field", and requiring the tori to be transverse to $\xi$;  this means  that $\xi$ is nowhere tangent to the torus.

In \cite{KMM23}, the Converse KAM criterion consisted in pushing forward the direction field $\xi_x$ from a point $x$ along $B$ (in  one direction of time) and testing whether it becomes a negative multiple of $\xi$ modulo $B$ at some time (i.e.~$\exists\, t,\mu \in \R, \lambda<0,$ such that $\phi_{t*}\xi_x = (\lambda \xi + \mu B)_{\phi_t x}$, where $\phi$ is the fieldline flow and $\phi_{t*}$ is its derivative at time $t$). 
In the present work, we implement a different yet equivalent approach, which we term the `conefield formulation'. 
In this formulation, we pull back a 1-form (in both directions of time) to produce upper and lower bounds on the ``slope" of invariant tori through any point; thus, for each point tested, we construct  a cone of possible slopes, forming a ``conefield". 
If at some stage the upper bound is below the lower bound then {the set of possible slopes is empty, so} there is no invariant torus through that point.

\subsection{Direction  field}
Following \cite{KMM23}, the Converse KAM method implemented here requires the choice of a continuous non-zero vector field $\xi$ to allow us to specify a class of tori of interest. The field $\xi$ can be considered a tangent vector to the 1D foliation required by the Converse KAM method of \cite{mackay2018finding} to identify regions through which pass no invariant tori of a vector field $B$ transverse to the foliation. 
Because only the direction matters, not the magnitude, we call $\xi$ a {\em direction field}. 
{For the principal class of tori, we chose $\xi$ to be in the direction of $\partial_\psi$.}

One more ingredient we take from \cite{KMM23} is the observation that for any positive function $f$, the vector field $V=B/f$ has the same invariant tori as $B$.  So it is convenient to choose a function $f$ to simplify the expression for $V$.  In general, $V$ no longer preserves the same volume form $\Omega$ as $B$, but preserves the related volume form $f\Omega$.  Also the important relation $i_B\beta = 0$ is inherited by $V$:~$i_V\beta=0$.
We will treat $B$ in {the rest of this section}, but one should bear in mind this possibly useful pre-processing {and we shall use it in the next section}.

\subsection{Conefield construction and nonexistence condition}
\label{subsec:conefields}

We describe here {the idea and construction of} conefields for the class of invariant tori of a 3D volume-preserving vector field $B$ that are transverse to a direction field $\xi$, by adapting the construction for area-preserving twist maps in \cite{mackay1985converse}. See also the brief description in \cite{mackay2018finding}.

If there is a (differentiable) invariant 2-torus $\mathcal{T}$ for $B$ then one can represent the tangent plane $T_x\mathcal{T}$ at a point $x \in \mathcal{T}$ as the kernel of a non-zero covector $\alpha_x$,
\begin{equation*}   
  T_x\mathcal{T} = \ker \alpha_x =  \{ v\in T_x M ; \alpha_x(v) = 0\} \,.
\end{equation*} 
In ordinary speak, $\alpha_x(v) = n_x \cdot v$ where $n_x$ is a normal vector to the tangent plane at $x$; but there is no need to invoke a Riemannian metric so we do not express it that way. A covector is a linear functional on a vector space and the set of covectors at a point of a manifold $M$ is called the cotangent space $T^*M$.

Multiplication of $\alpha_x$ by any non-zero scalar produces the same kernel, so one could consider $\alpha_x$ to be in the quotient by non-zero scalars of the cotangent space minus the origin.  It is convenient, however, to assign a ``co-orientation'' to a 2D subspace in 3D, i.e.~to say one side of it is positive and the other negative.  This can be done by the sign of $\alpha_x$ off its kernel.  So we consider $\alpha_x$ to be in the quotient of $T_x^*M \setminus \{0\}$ by only positive scalars.

Consistently using the subscript $_x$ to indicate evaluation at a point $x$ will lead to excessive notation when we  later introduce time-dependent covectors and add superscripts $\pm$. Therefore, we will omit the subscript when the context makes it clear.

For a torus transverse to $\xi$, such a covector satisfies $\alpha(\xi) \neq 0$. 
A natural co-orientation is induced by taking $\alpha(\xi)>0$, corresponding to saying $\xi$ is on the positive side of $\ker \alpha$.

The tangent plane to an invariant torus at a point $x$ contains the vector $B_x$, so we can restrict attention to covectors $\alpha$ such that $\alpha(B)=0$.

For a covector $\alpha$ satisfying both $\alpha(\xi)>0$ and $\alpha(B)=0$, we can define its ``slope'', corresponding to a notion of slope of the plane defined by its kernel.  To do this, we introduce another reference vector field $\eta$.
Let $\Omega$ be the volume-form in $M$ (that is preserved by $B$) and 
let $\eta$ be a vector field such that $(\eta,\xi,B)$ form a positively oriented frame, i.e.~$\Omega(\eta,\xi,B)>0$. 
For example, one can take $\eta = \xi \times B$ (though when we choose a transverse section to $B$ later, it will be convenient to choose both $\xi$ and $\eta$ to be tangent to the section by subtracting off suitable multiples of $B$). We call $\eta$ a {\em horizontal field}.
Then we define the \emph{slope} $\sigma \in \R$ of such a covector $\alpha$ by
\begin{equation}
\sigma = \frac{\alpha(\eta)}{\alpha(\xi)} \,.
\end{equation}
The justification for the name is that $\alpha(\eta+\sigma \xi)=0$ so $\eta+\sigma \xi \in \ker \alpha$; this vector has slope $\sigma$ in coordinates with horizontal unit vector $-\eta$ and vertical unit vector $\xi$ (perhaps it would be considered more natural to take the opposite sign for $\eta$ and define $\sigma$ with a minus sign).
We say one such covector $\alpha^+$ is greater than or equal to another $\alpha^-$ iff  its slope is greater than or equal to the slope of the other; we write this as $\alpha^+ \succeq \alpha^-$.
The same definition can be applied to covectors with $\alpha(\xi)<0, \alpha(B)=0$, but it is important to realise that if $\alpha^+\succeq \alpha^-$ then also $-\alpha^+\succeq -\alpha^-$.

We extend the definition of the order on covectors $\alpha$ with $\alpha(B)=0$ to the multiples of \begin{equation}
\alpha_0 = i_Bi_\xi\Omega
\label{eq:alpha0}
\end{equation}
(those covectors for which $\alpha(\xi)=0$), by defining $\alpha_0 \succ \alpha \succ -\alpha_0$ for all $\alpha$ with $\alpha(\xi)>0$ and we define the corresponding slopes to be $\pm \infty$.

\begin{figure}[htb]
\centering
\includegraphics[height=2.0in]{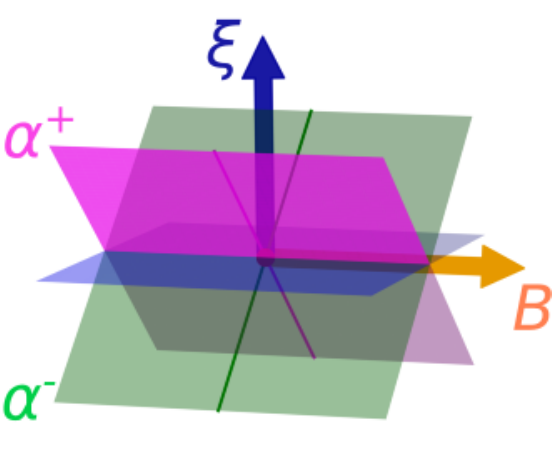}
\caption{A cone, consisting of all the covectors at a point whose kernels are planes, like the blue one, that lie in the sector between those for  $ \alpha^+$  and $\alpha^-$  that does not contain $\xi$.} 
\label{fig:cone}
\end{figure}

The conefield method constructs bounds on the slopes of all invariant tori that are transverse to $\xi$.  A {\em cone} at a point $x \in M$ is a set $C_x$ of non-zero covectors at $x$; in practice, we will take cones of the form $\alpha_x^- \preceq \alpha_x \preceq \alpha_x^+$ for some $\alpha_x^\pm$ (see Figure~\ref{fig:cone}).  
A {\em conefield} $C$ is a choice of cone at each point of $M$ (to allow the killends extension, we will require $C_x$ to be measurable with respect to $x$).  We say a conefield {\em bounds} the slopes of all invariant tori transverse to $\xi$ if for all points $x$ on such an invariant torus the cone $C_x$ contains a covector representing the tangent plane to the torus there.

We will start with the conefield $C$ defined by $\alpha^+ = \alpha_0, \alpha^- = -\alpha_0$.  It trivially bounds the slopes of all invariant tori of the given class, because it contains all possible slopes at all points. Indeed, by transversality, we have the strict order $-\alpha_0 \prec \alpha \prec \alpha_0$ for any covector $\alpha$ representing the tangent plane to an invariant torus of the given class. But after an initial stage of an iteration that we will describe, we will typically obtain that $\alpha^\pm(\xi)>0$ at all points.

Now we will  modify $\alpha^{\pm}$ iteratively such that if there is an invariant torus $\mathcal{T}$ with {tangent plane given by}  $\ker \alpha$ at the given point then at all stages $\alpha^- \prec \alpha$ and $\alpha \prec \alpha^+$. That is, $\alpha^\pm$ provide (strict) upper and lower bounds on the slope of any invariant torus of the given class through the given point. If at any stage our construction gives $\alpha^- \succeq \alpha^+$, then we deduce that no such $\alpha$ exists, and hence, there is no invariant torus of the given class through the given point. This will give our primary non-existence condition.

Denote the flow of $B$ for time $t\in \mathbb{R}$ by $\phi_t$ (i.e.~$\phi_t(x_0)$ is the solution of $\dot{x}{(t)} = B(x(t))$ from $x(0) = x_0$) and its pullback by $\phi_{t}^*$.
%\vspace{0.3cm}

For given choice of direction field $\xi$, let $\alpha_0$ be the 1-form defined in (\ref{eq:alpha0}), 
{equivalently, $\alpha_0 = (\xi \times B)^\flat$,}
and note that $\xi$ transverse to $\mathcal{T}$ implies $\alpha_0 \neq 0$ at all points of $\mathcal{T}$ ($\alpha_0 \ne 0$ means there exists a vector $v$ on which $\alpha_0(v) \ne 0$, e.g.~$v = \xi \times B$). Then compute the pullback 
$${\alpha}_t = \phi_t^*(\alpha_0)$$ for $t\in [-\tau,\tau]$, some $\tau > 0$. This can be done by integrating
the adjoint linearised equation about the field line, but it is equivalent to compute the matrix solution of the linearised fieldline flow and apply that to the covector (as we will spell out below).

Note that $\alpha_t (B) = 0$ for all $t$. Also, $\alpha_0(\xi) = 0$. If $\alpha_t(\xi)$ happens to be zero for all $t$ then we will deduce nothing, but that is a very special (“shearless”) situation. 
Let $T_{\pm}$ be the sets {of times $t$} for which $\alpha_t(\xi)$ is positive and negative, respectively. Let $\sigma^+$ be the minimum  of the slopes of $\alpha_t$  for $t\in T_+$,  and $\sigma^-$ be the maximum of the slopes of $-\alpha_t$ for $t\in T_-$ (the slope of $-\alpha_t$ is the same as for $\alpha_t$ but we include the minus sign because the resulting $\alpha^-$ that we want to use is $-\alpha_t$ for the maximising $t \in T_-$). 
If there is an invariant torus of a given class through the given point, then its slope $\sigma$ satisfies
\begin{equation*}
\sigma^- <  \sigma  < \sigma^+ \,.
\end{equation*}
We define the resulting cone by $\alpha^+ = \alpha_t$ for the minimising value of $t \in T_+$ and $\alpha^- = -\alpha_t$ for the maximising value of $t \in T_-$.

As $\tau$ is increased, the cone at a point can only get tighter.
If the case $\sigma^+ \leq \sigma^-$  {occurs} during the computation, it will imply that such an invariant torus $\mathcal{T}$ does not exist.
This is a reformulation of the Converse KAM method described in \cite{KMM23}. The advantage of the reformulation is that for points that are not eliminated we still get some information, namely bounds on the slopes of possible invariant tori through them.

The sets $T_\pm$ depend on the ``shear" of the field $B$ relative to $\xi$. We say $B$ has {\it positive shear} relative to $\xi$ if $\frac{d}{dt} (\alpha_t\xi) < 0$ at $t=0$ and negative shear if $\frac{d}{dt} (\alpha_t\xi) > 0$ (this looks backwards, but $\alpha_t$ is the pullback of $\alpha_0$). 
In cases of positive shear, then $\alpha_t(\xi) > 0$ for all small negative $t$, and negative for all small positive $t$. Furthermore, if the shear remains positive along the fieldline then the slope of $\alpha_t$ decreases from $+\infty$ as $t$ decreases from $0$, as long as $\alpha_t(\xi)$ remains positive.
Similarly, the slope of $-\alpha_t$ increases from $-\infty$ as $t$ increases from $0$, as long as $-\alpha_t(\xi)$ remains positive. So in this case the set $T_+$ consists of negative times and the set $T_-$ consists of positive times. 
Denoting $t_+$ the backwards integration time and $t_-$ the forwards integration time, 
if either of $\pm\alpha_{t_\pm} (\xi)$ reaches $0$ as $t_+$ decreases or $t_-$ increases, then the slopes will already have failed to satisfy $\sigma^- < \sigma^+$ and one can stop integrating. 
So, in the case of positive shear the maximum or minimum slopes are achieved for the longest (backwards and forwards, respectively) times computed, simplifying the computation. The same applies with obvious changes of sign for negative shear. An advantage of the general procedure above, however, is that it can be applied to cases without definite shear. The shear might be positive in some regions, negative in others, and the fieldline might move from one to the other.

%% THEOREM % % 
 {Summarising}, the conefield formulation translated to magnetic fields  says that
\begin{thm}\label{thm:CF}
{For a magnetic field $B$, direction field $\xi$ and horizontal field $\eta$,} given a point $x_0=x(0)$ and time $\tau>0$, let $\alpha_t$ be the pullback of the covector $\alpha_{0}=i_B i_\xi\Omega$ from $x(t)$ to $x_0$ along the fieldline flow for all $t 
 \in [-\tau,\tau]$ and compute the resulting slope $\sigma_t$. %
If there are times $t_\pm$ such that $\sigma_{t^+}  \le \sigma_{t^-}$ and $\pm \alpha_{t_\pm}(\xi) > 0$, then there is no invariant torus for $B$ through $x_0$ transverse to $\xi$. 
\end{thm}
In this statement, we didn't insist on finding the maximum and minimum of $\sigma_t$ over $t \in T_\pm$, because the extra freedom is useful in practice.

%\vspace{0.5cm}
To make it more explicit, the method consists of:
\begin{enumerate}
    \item From initial condition: $x(0)=x_0$, $M(0)= I$ (the $3\times 3$ identity matrix), solve 
     \begin{subequations}
     \begin{align}
        \dot{x}(t) &= B(x(t)) \,,\\
        \dot{M}(t) &= DB_{x(t)}\, M{(t)}\,,
     \end{align}
     \end{subequations}
     for $t\in(-\tau,\tau)$ for increasing $\tau$.
     
    \item  Compute %
    $\alpha_t (\xi)$ and $\alpha_t (\eta)$, using that for any vector $w$ at $x_0$, $\alpha_t (w) = \alpha_0\, M(t)\, w = \Omega(\xi,B, M(t)w)$, where $\Omega$, $\xi$ and $B$ are computed at $x(t)$. 
    
    \item Find the sets $T_\pm$ 
    of $t$ where $\alpha_t (\xi)$  is computed to be positive/negative at the chosen $x_0$.
    
    \item Find 
    the minimum, $\sigma^+$, of $\alpha_t(\eta)/\alpha_t(\xi)$ over $t\in T_+$, and the maximum, $\sigma^-$, of $\alpha_t(\eta)/\alpha_t(\xi)$ over $t\in T_{-}$. 
    
    \item {Compare $\sigma^+$ and $\sigma^-$. If $\sigma^- < \sigma^+$, increase $\tau$ and recompute until $\sigma^+ < \sigma^-$ or $\tau$ reaches a predetermined timeout.}
    
    \item Define $\alpha_\tau^+$ to be $\alpha_t$ at the maximising time $t \in T_+$ and $\alpha_\tau^-$ to be $-\alpha_t$ at the minimising time $t \in T_-$, and $\sigma^\pm_\tau$ the corresponding slopes.
\end{enumerate}

\begin{figure}[h!]
 \centering  
   \includegraphics[height=9.1cm]{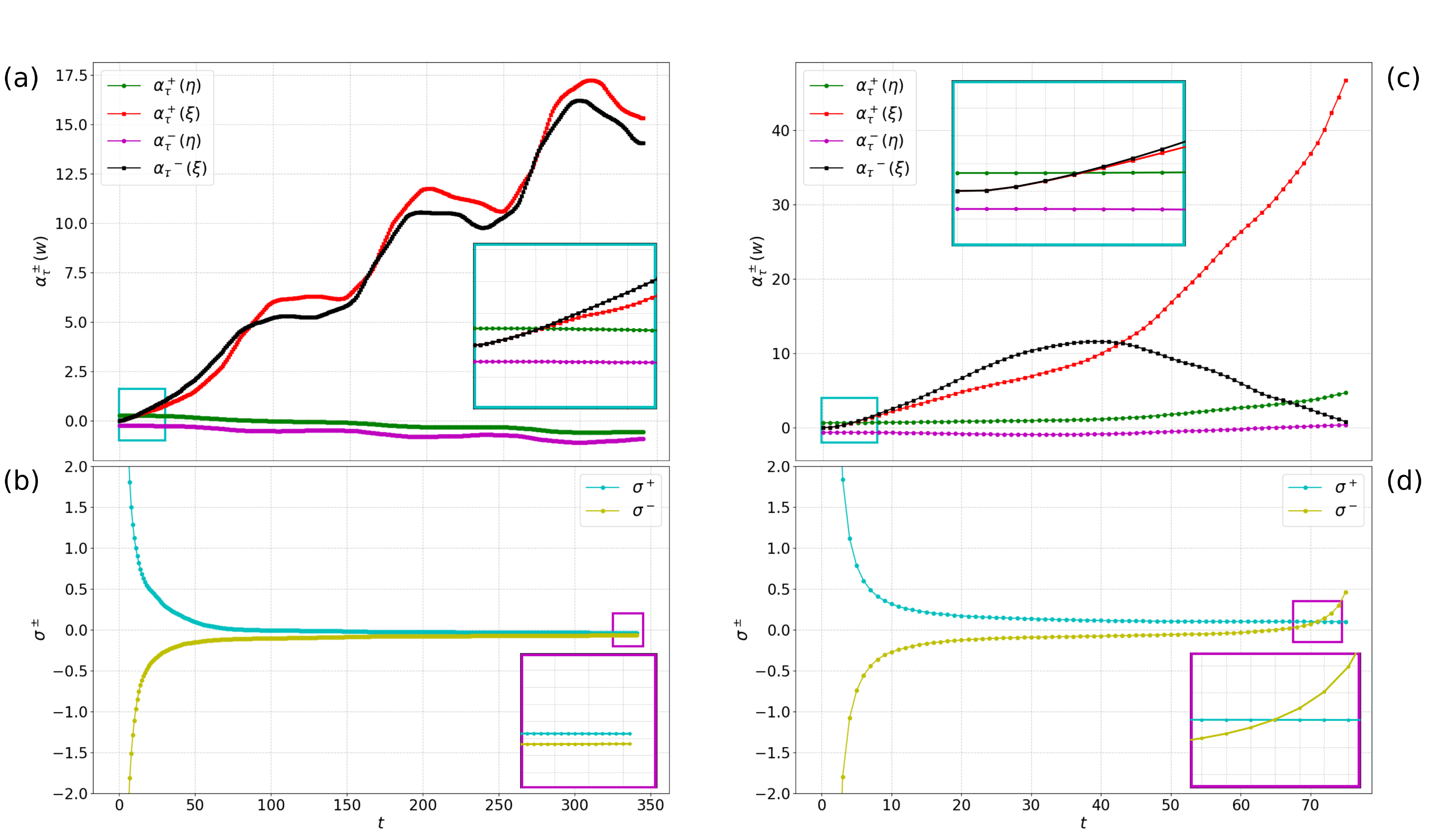}
   \caption{Typical evolution of the values of {$\alpha_\tau^\pm(\eta)$ and $ \alpha_\tau^\pm(\xi)$} (top), along with the associated slopes {$\sigma_\tau^\pm$} (bottom) for the magnetic field derived from (\ref{A_ex2}), for two different initial conditions: [(a)(b)] a point in the complement of the non-existence region ($(\tilde{y}_1,\tilde{z}_1) = (0.3,0.2)$); and [(c)(d)] a point inside the non-existence region ($(\tilde{y}_2,\tilde{z}_2) = (0.35,0.45)$). The insets show blowups of the indicated regions.}
   \label{Fig_conefield_example}
\end{figure}

Figure \ref{Fig_conefield_example} illustrates the typical evolution
of {$\alpha_\tau^\pm(\eta)$ and $\alpha_\tau^\pm(\xi)$}, along with the associated slopes {$\sigma_\tau^\pm$}, as functions of time {$\tau$} for {two} initial conditions for the example magnetic fields that we study here,\footnote{The specific parameter values of the two simulations correspond to (\ref{eq:standard_values}), (\ref{A_ex2}) with $\varepsilon_{21} = \varepsilon_{32} = 0.003$.} which have positive shear $(w_2>0)$.
The left {column}, corresponding to a point outside the non-existence region, shows how the cone at the point becomes narrower as time evolves, with the slopes approaching a common value. The right {column}, corresponding to a point in the non-existence region, shows how the computation is stopped when {$\sigma_\tau^+ < \sigma_\tau^-$ (to make the crossing clear, we integrated a bit further than necessary)}.
Note that the integration time is much shorter for the second case.
Note also that on an orbit with positive Lyapunov exponent, $\alpha_t$ can be expected to grow at the Lyapunov rate, but the power of the Converse KAM method is that typically, the greater the Lyapunov exponent, the shorter the time it takes to reach a decision.

%////////////////////////////
%////////////////////////////

For use in the next subsection, it is convenient to transform the slope $\sigma$, defined relative to the direction field $\xi$ and  {horizontal field} $\eta$, into a slope $m$ in a reference coordinate system (e.g., $(R, z)$ for a poloidal plane $\phi = \phi_0$). Assuming that {$\xi$ and} $\eta$ are tangent to the chosen poloidal plane, we can express the projections of $\alpha$ in components as $\alpha(\xi) = \alpha_R \xi^R + \alpha_z \xi^z$ and $\alpha(\eta) = \alpha_R \eta^R + \alpha_z \eta^z$, and solve for $\{\alpha_R, \alpha_z\}$, as $\xi$ and $\eta$ are linearly independent by construction. 
Recalling that $(\alpha_R, \alpha_z)$ are the covariant components of a {co}vector, the transformed slope is simply given by $m = -{\alpha_R}/{\alpha_z}$, that is,
\begin{equation}
\label{Eq_m}
m = \frac{dz}{dR}= \frac{\sigma\xi^z - \eta^z}{\sigma\xi^R - \eta^R} \,.
\end{equation}

\subsection{Killends}
\label{subsec:killends}
\begin{figure}[h!]
 \centering  
    \includegraphics[height=3.5cm]{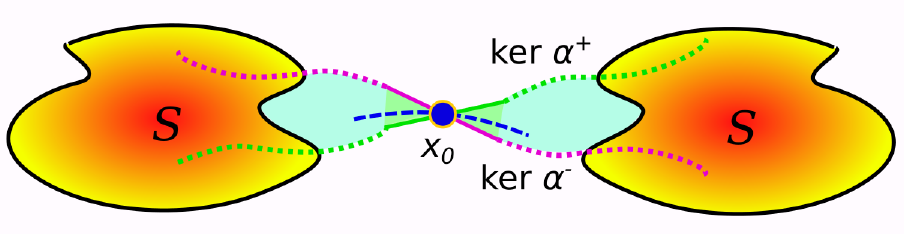} 
   \caption{Diagram of a transverse section in which the conefield data allows identification of a point (blue) where any possible torus passing through it (e.g.~the blue dashed curve) enters a region $\mathcal{S}$ where non-existence has been established, thus allowing us to add it to the non-existence region.}
   \label{Fig_killends1}
\end{figure}

As mentioned in the introduction, %Sec.~\ref{sec:intr}, 
the conefield information for points that are not eliminated by the method in Sec.~\ref{subsec:conefields} can be used to extend the region of non-existence. The approach described here follows a similar rationale to that in \cite{mackay1985converse}, where the term ``killends" was introduced to describe a subsidiary criterion that allows the extension of a non-existence region previously computed. The general idea is to identify any point outside the non-existence region for which the integration of {the differential inclusion defined by the conefield}  ensures that any possible torus passing through the point enters the previously computed non-existence region. This allows us to add the point and its orbit to an enlarged non-existence region (see figure \ref{Fig_killends1}).

%////////////////////////////
%////////////////////////////

Suppose one has established a set $\mathcal{S}$ of points through which there pass no invariant
torus transverse to $\xi$ and two fields of covectors $\alpha^-$, $\alpha^+$ on the complement of $\mathcal{S}$ such that the tangent plane represented by $\alpha$ to any invariant torus satisfies $\alpha^- \preceq \alpha \preceq \alpha^+$. To simplify the presentation, let us assume that $\alpha^\pm(\xi) > 0$, as above.

Define the {\it right cone} at $x \in M$ to be the set of tangent vectors $v$ such that $\alpha^-(v) \geq 0$, $\alpha^+ (v) \leq 0$. The tangents $v$ to an invariant torus with $\alpha(v) > 0$ must lie in the right cone. Choose a surface transverse to $B$. If all differentiable paths from $x$ with tangent in the right cone at each point reach $\mathcal{S}$, then $x$ can be added to $\mathcal{S}$, because any invariant torus through $x$ would contain such a path and hence intersect $\mathcal{S}$. We can define in analogous manner the {\it left cone} and study the opposite direction, however the analysis will yield the same result.  

\begin{figure}[h!]
 \centering  
    \includegraphics[height=3.5cm]{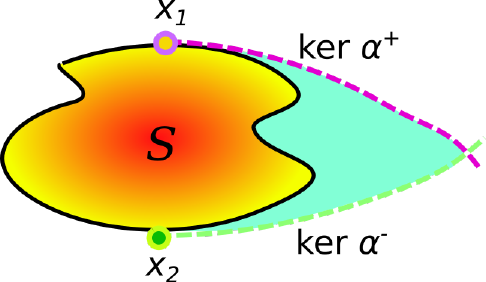} 
   \caption{Diagram of the alternative `killends' approach to enlarge the non-existence region $\mathcal{S}$ by considering the continuation of the cones of points tangent to $\mathcal{S}$.}
   \label{Fig_killends2}
\end{figure}

A more efficient method is to start from the boundary of $\mathcal{S}$ and enlarge $\mathcal{S}$ using
the bounds on allowed slopes, as follows. For simplicity of presentation, suppose the
boundary of $\mathcal{S}$ is differentiable. Take two points $x_1, x_2$ of $\partial \mathcal{S}$ connected by an arc of $\partial \mathcal{S}$ such that $\mathrm{ker}\, \alpha^+$ is tangent to $\partial \mathcal{S}$ at $x_1$ and $\mathrm{ker}\, \alpha^-$ is tangent to $\partial \mathcal{S}$ at $x_2$, see figure \ref{Fig_killends2}. 
Integrate $\mathrm{ker}\, \alpha^\pm$ leftwards in the chosen section until they intersect. The region they bound with $\partial \mathcal{S}$ can be added to $\mathcal{S}$. The similar procedure applies at the right of $\mathcal{S}$.
In the present work, we will refer to the curves created by the integration of $\mathrm{ker}\, \alpha^\pm$ as `killends branches'.
%

%////////////////////////////
%////////////////////////////

In an ideal situation, the branches computed by the integration of $\mathrm{ker}\, \alpha^\pm$ from two points $x_1, x_2 \in \partial \mathcal{S}$ will intersect away from the boundary of $\mathcal{S}$, allowing us to add all the area between the two branches to $\mathcal{S}$, as shown in figure \ref{Fig_killends2}. However, in practice, we cannot guarantee that $\partial \mathcal{S}$ is differentiable, nor that a point chosen on the boundary of the computed $\mathcal{S}$ is an exact point on the boundary of the non-existence region. In our examples, we compute an approximate $\tilde{\mathcal{S}}$ from a regular grid of initial conditions on a poloidal section, as the points eliminated by the Converse KAM method for a predetermined time-out $t_f$.

The points on the grid neighbouring $\tilde{\mathcal{S}}$ may at best lie outside the analytic $\mathcal{S}$ and at worst may be points that would be eliminated by the method using a larger time-out. Because of this, the integration of $\mathrm{ker}\, \alpha^\pm$ from approximate data needs to consider the scenario in which the continuation branch $\mathrm{ker}\, \alpha^\pm$, starting from a given $x_0 \in \partial\tilde{\mathcal{S}}$, enters the non-existence region. In such a case, the computed branch is added to $\tilde{\mathcal{S}}$, and a different starting point $x_1$ must be chosen to start the process again in order to produce intersecting branches, as depicted in figure \ref{Fig_killends2}.

In our implementation, the integration of $\mathrm{ker}\, \alpha^\pm$ starting from a point $x_0 \in \partial \mathcal{S}$ is computed by taking Euler steps with the associated slope and recomputing slopes for each new point. To be precise, on the poloidal plane and in symplectic polar coordinates, the points $p_j$ on a branch of $\mathrm{ker}\, \alpha^\pm$ are calculated by
\begin{equation}
    p_{k+1} = p_k + \Delta\, (1,m^{\pm}(p_k))^T\,, \qquad p_0 = x_0\,,
\end{equation}
with $m^{\pm}$ from (\ref{Eq_m}) and an a priori $\Delta > 0$.
We choose seed points $x_i \in \partial \tilde{\mathcal{S}}$ to compute killends branches 
$\mathrm{ker}\, \alpha^\pm$, with $x_i$ being the closest point on the boundary to the estimated position of the O-points of the resonances considered. 
If, by any chance, the integration of the kernel enters the non-existence region at the integration step $k+1$, we found as the most convenient strategy to select a new seed point by integrating $p_k$ with the opposite slope for one step. For example, we define $y_i$ as: $y_i = p_k + \Delta\, (1,m^{\mp}(p_k))^T$, and then proceed to integrate forward and backward.

Note that, by construction, the area enclosed by the  
killends branches  
provides {a lower %upper 
bound} on the full region of non-existence of invariant tori. 
As will be shown in the next section, however, the estimates include regions that are typically difficult (even impossible) to detect efficiently with the basic Converse KAM  method, such as neighbourhoods of cantori and principal hyperbolic periodic orbits.
Following \cite{stark1988exhaustive}, 
it is expected that the region eliminated by the conefield and killends method can be made arbitrarily close to the full non-existence region with sufficient computational work.
The numerical estimation may be affected, however, by seed points that are only approximately close to the boundary of $\mathcal{S}$.

\section{Results}
\label{sec:results}

In this section, we apply the conefield formulation of the Converse KAM method along with the killends subsidiary criterion discussed in Section \ref{sec:cKAM}, to non-integrable magnetic fields of the type introduced in Section \ref{sec:magfields}. Specifically, we use the method to identify regions without invariant tori (i.e., flux surfaces) transverse to the $\psi$-direction. These regions, which do not lie on such tori,  {correspond to}  magnetic islands and chaotic zones.

In detail, the method is applied to regular grids of initial conditions in symplectic coordinates $(\tilde{y},\tilde{z})$, %(\ref{eq:sym_cords}) 
\begin{align}
\label{eq:sym_cords}
\begin{split}
\tilde{y}& = \sqrt{2\psi/B_0}\cos\vartheta \,, \\
\tilde{z}& = \sqrt{2\psi/B_0}\sin{\vartheta} \,,
\end{split}
\end{align}
over the plane $\phi=0$, for the magnetic fields defined by (\ref{eq:fields}). The grid resolution is $160\times 160$ initial conditions for each sample. By counting the initial conditions identified by the method, we can give a lower bound on the area (which corresponds to the toroidal flux in symplectic coordinates) that is not occupied by tori transverse to the specified direction field.

It is important to highlight that areas are equivalent whether computed in symplectic coordinates $(\tilde{y}, \tilde{z})$ or in $(\psi, \vartheta)$, since $d\tilde{y} \wedge d\tilde{z} = B_0^{-1} d\psi \wedge d\vartheta$. Consequently, the area $S$ of the non-existence region in the plane $\phi=0$ can be estimated by counting the number of initial conditions detected by the method on a regular grid in $(\tilde{y}, \tilde{z}, \phi=0)$. In other words, if $\mathcal{S}$ represents the set of points identified by the conefield formulation on an $N \times N$ regular grid over the domain $[\tilde{y}_0 - L, \tilde{y}_0 + L] \times [\tilde{z}_0 - L, \tilde{z}_0 + L]$, the area $S$ is approximated by
\begin{equation}
    \label{eq:area}
    S \sim \frac{4L^2}{N^2}\sum_{i=1}^N \sum_{j=1}^N 1_{\{(\tilde{y}_i,\Tilde{z}_j)\in \mathcal{S}\}}\, .
\end{equation}

In order to account for the possibility that a trajectory does not fulfill the termination condition, we set a timeout $t_f$; thus the interval of integration for the method is $[-t_f/2,t_f/2]$. If this timeout is reached, the status of the corresponding initial condition remains undecided. Naturally, this category includes all initial conditions lying on invariant tori of the specified class, but it may also include others that require more time for the nonexistence to be detected. The timeout values do not necessarily reflect the length of the trajectories.

For a given trajectory, the interval of integration is reported as the time $t_*$ at which non-existence was detected, or $t_f$ if it was not detected. As a measure of non-existence of tori of a given class, figures display the relative time of detection shown in hues, using the ratio $q = t_*/t_{f}$, which represents the time of detection relative to the timeout.

In all the examples throughout this section and elsewhere, we take the following values and functions for the vector potential (\ref{eq:potential})
\begin{align}
\label{eq:standard_values}
\begin{split}
\begin{aligned}
w_1& = 1/4,\\
w_2& = 1,
\end{aligned}\qquad
\begin{aligned}
B_0&=1,\\
R_0&=2,
\end{aligned}\qquad
\begin{aligned}
\zeta_{mn}&= 0,\\
f_{mn}(\psi)&= \psi-R_0^2/B_0.
\end{aligned}
\end{split}
\end{align}
We select only finitely many of the $\varepsilon_{mn}$ to be non-zero.
As previously mentioned, the results in all  forthcoming figures are presented over the poloidal plane $\phi=0$.

\subsection{Conefield for a non-integrable example}
\label{subsec:res_conefields}
We consider magnetic fields with two helical terms, derived from (\ref{eq:potential}). We omit the integrable case with only one helical term as it was already considered extensively in \cite{KMM23} and the present method is expected to yield the same results. 
With the choice of a radial direction field ($\xi = \partial_\psi$), the method is able to identify and eliminate points (and their field lines) that do not lie on tori of the original class. 
The method, like {that} in \cite{KMM23}, does not distinguish between points lying on tori of another class or in chaotic regions. However, if needed, employing a suitable foliation centered on the elliptic field lines of an island chain could differentiate between these two cases.

The example considered corresponds to the magnetic field derived from (\ref{eq:potential}) for two modes now, namely the resonances $2/1$ and $3/2$ with same perturbation parameter value $\eps_{21} = \eps_{32} = \eps$, that is
\begin{equation}
\label{A_ex2}
A_\phi = -\left[\psi/4 + \psi^2 + \eps\psi(\psi-4)\left[\cos(2\vartheta-\phi) + \psi^{1/2}\cos(3\vartheta-2\phi)\right]\right].
\end{equation}

\begin{figure}[h!]
 \centering  
   \includegraphics[height=7.1cm, trim={20 0 0 0},clip]{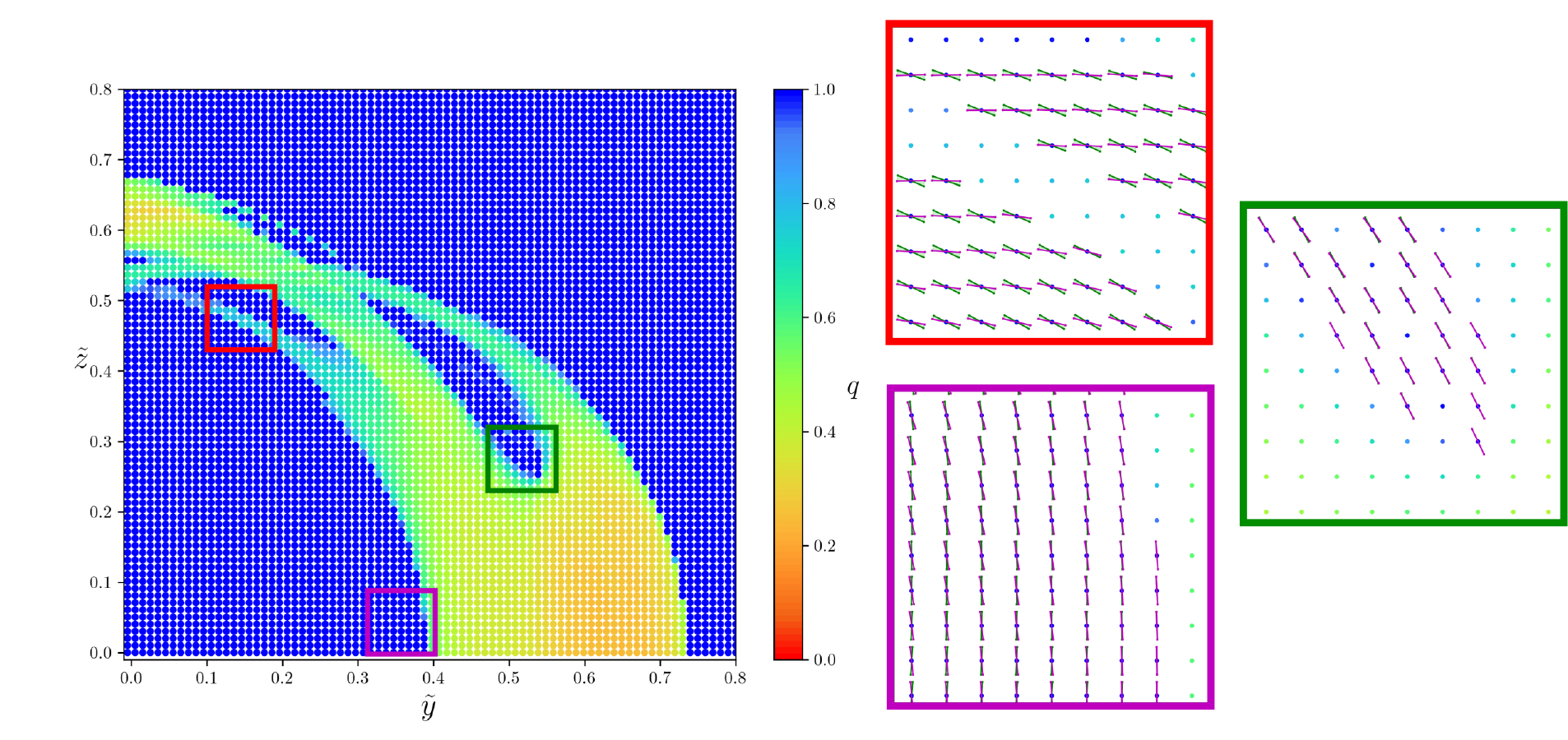}
   \caption{Conefield results (shown in hues) and three zoomed-in insets (on the left) displaying the computed conefield on the complement of the detected non-existence region $\tilde{\mathcal{S}}$, on a regular grid of initial conditions in the positive quadrant of the poloidal plane $\phi = 0$ for the magnetic field derived from (\ref{A_ex2}) with $\varepsilon = 0.003$ in symplectic coordinates. The hues vary from fast detection (red) to no detection at all (blue) within the timeout $t_f = 80$ (which corresponds to about 15 toroidal revolutions). $q$ denotes the ratio of the time to detection to the timeout, except that $q=1$ denotes no detection within the timeout.}
   \label{Fig_conefield_complement}
\end{figure}

%Figure \ref{Fig_conefield_boundary} [{\it or} 
Figure \ref{Fig_conefield_complement} 
shows the results of applying the conefield formulation to the positive quadrant of the poloidal plane in symplectic coordinates ($\tilde{y} > 0, \tilde{z} > 0$).
Points in colours other than blue are points through which the method detects that there are no invariant tori of the desired class because the cone becomes empty at some time. The hues vary from fast detection (red) to slower detection (near blue).  The blue points remain undecided within the timeout $t_f = 80$. The figure includes three additional zoomed-in insets on the right, which illustrate the conefield as computed by the method on the 
complement of $\tilde{\mathcal{S}}$, i.e.~blue points).

%////////////////////////////
%////////////////////////////

\begin{figure}[h!]
 \centering  
   \includegraphics[height=7.4cm, trim={50 0 70 0},clip]{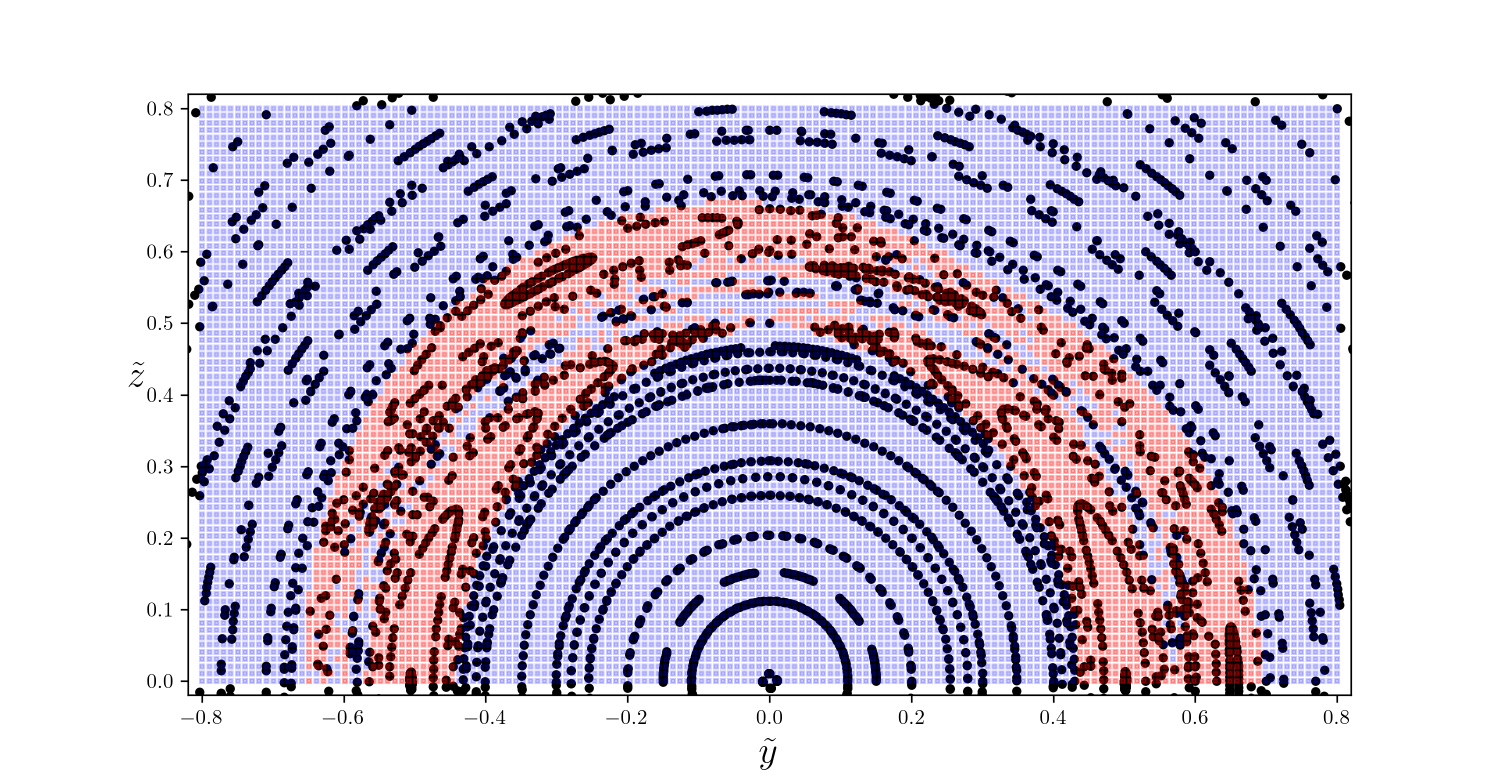} 
   \caption{Superposition of Poincaré plot (orbits in black) %green and dark blue) 
   and conefield results (red for non-existence, blue for the rest of area tested) within timeout $t_f = 300$
   on a regular grid of initial conditions in the upper half of the poloidal plane $\phi = 0$ for the magnetic field derived from (\ref{A_ex2}) with $\varepsilon = 0.001$ in symplectic coordinates.}
   \label{Fig_conefield_method}
\end{figure}

Figure \ref{Fig_conefield_method} shows a superposition of a Poincaré plot and the results of the conefield method  for the magnetic field derived from (\ref{A_ex2}) on the upper half of the poloidal plane $\phi = 0$ with $\varepsilon = 0.001$. {We see that the eliminated region corresponds to the main islands and chaotic regions.  The outer and inner regions that look close to integrable with tori transverse to the radial direction have not been eliminated, as indeed they should not.  There are also blue regions mixed with the red region that have not been eliminated but look as though they probably have no invariant torus of the given class through them.  Indeed, there are hyperbolic sets (of cantori and periodic orbits) that the conefield method per se can not eliminate.  But they can be eliminated by the killends extension, as will be illustrated in the next subsection.}

%////////////////////////////
%////////////////////////////

Before proceeding to application of the killends extension, we present results on how the area eliminated by the conefield method changes with the timeout, and we compare the results with those of what we call the direct method from \cite{KMM23}.

\begin{figure}[ht!]
 \centering   
   \includegraphics[height=8.2cm]{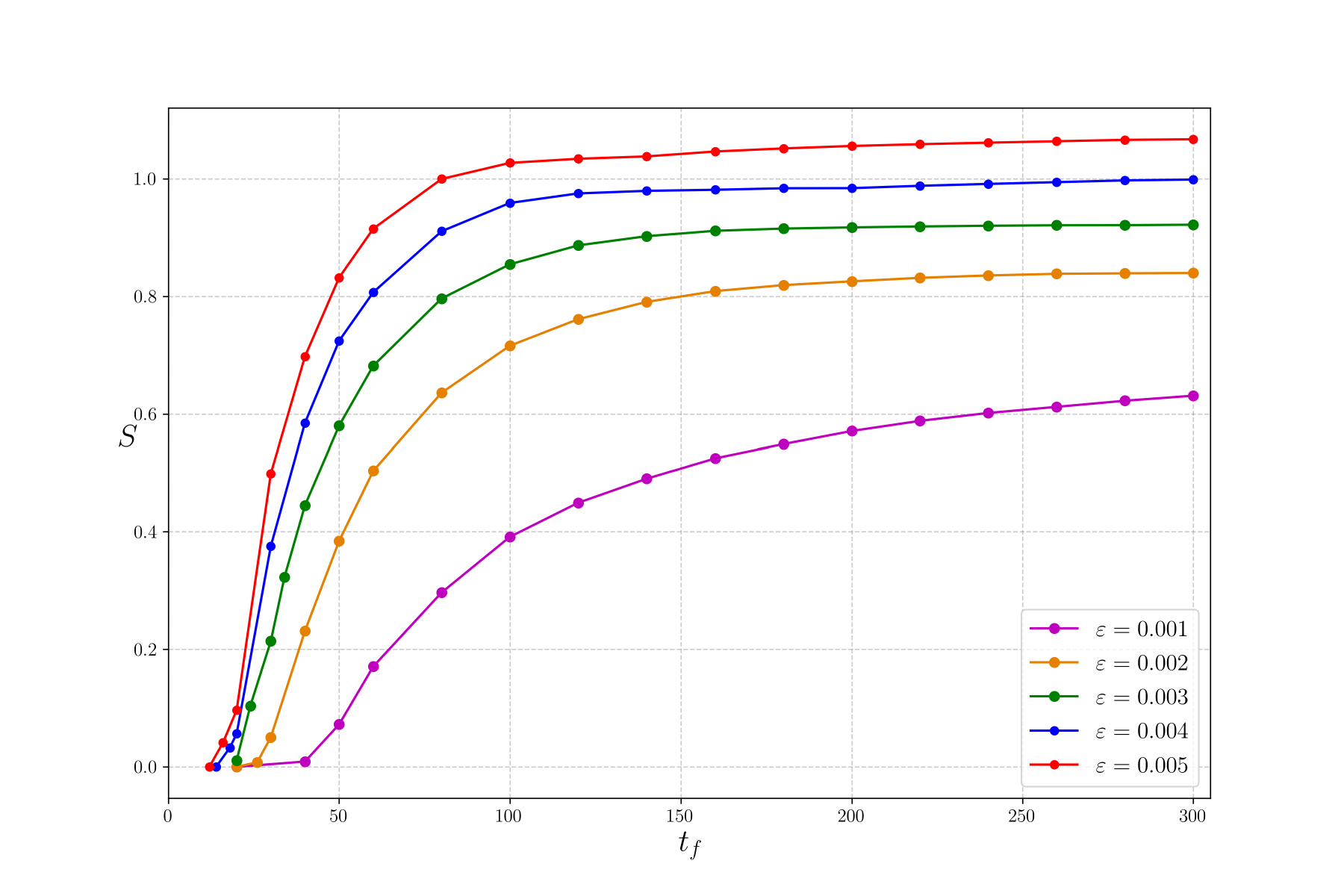}
   \caption{Nonexistence area $S(t_f)$ detected by Converse KAM conefield method for different values of $\eps$ for the field (\ref{A_ex2}).}
   \label{Fig_area_2res}
\end{figure}

Figure \ref{Fig_area_2res} shows the computed area $S=S(t_f)$ of nonexistence from the conefield approach for the present example for different values of the perturbation parameter $\eps$. We see that the estimated areas increase monotonically with $t_{f}$ and seem to be approaching a limit.
%////////////////////////////
%////////////////////////////

\begin{figure}[ht!]
 \centering   
    \includegraphics[height=8.2cm]{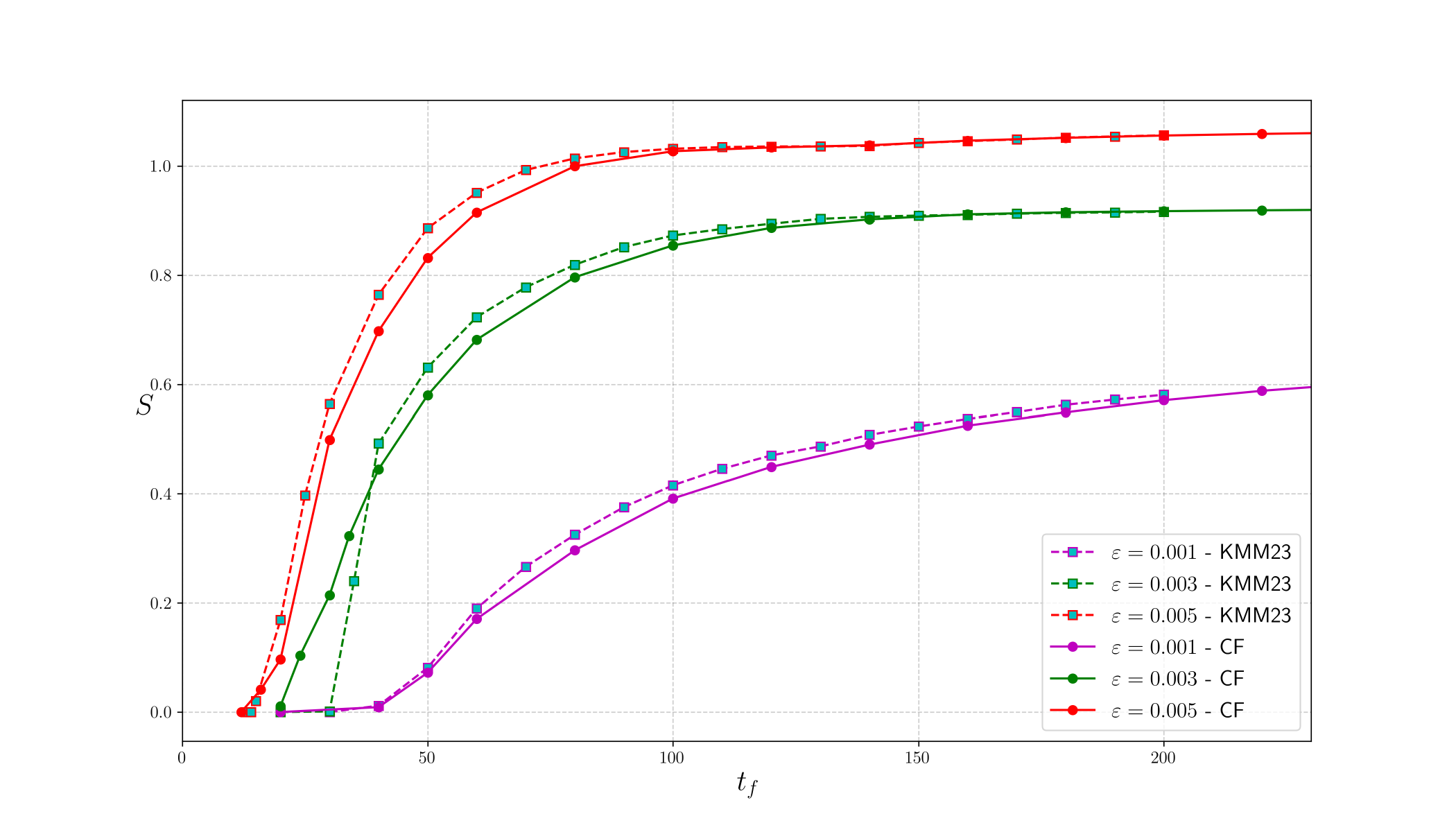}
   \caption{Nonexistence area $S(t_f)$ detected  for three different values of $\eps$ for the field (\ref{A_ex2}), by two Converse KAM methods: \emph{conefield approach} (solid line) and \emph{direct approach} in \cite{KMM23}.
}
   \label{Fig_paper_comparison}
\end{figure}

Figure \ref{Fig_paper_comparison} shows the computed area $S=S(t_f)$ of nonexistence from the conefield approach compared with the results in \cite{KMM23} for the same example for two different values of $\eps$. It can be seen that the two implementations yield similar results. In theory (compare \cite{mackay1985converse}), we would expect the results to yield almost identical plots for the same integration time, 
because the methods are equivalent except for the time-interval of integration:~$[0,t_f]$ for \cite{KMM23} versus $[-t_f/2,t_f/2]$ for the conefield method.  But as our initial conditions are only on $\phi=0$, this does mean that a slightly different sample of phase space is selected.
The difference in results may reflect this difference in the origin of the time-interval.
Regardless, the results are similar for a modest integration interval of $t_f = 300 \sim 48 $ laps around the $z$-axis.
%////////////////////////////
%////////////////////////////

\subsection{Killends in a non-integrable example}
\label{subsec:res_killends}
 
To construct the killends branches 
and measure the area enclosed by them on the poloidal plane $\phi = 0$ for the example field (\ref{A_ex2}), we took into account the expected shape of the magnetic islands for the chosen resonances to select suitable seed points for the computation. For exposition purposes, we restrict our attention to the upper half of the poloidal plane, as the example fields are symmetric with respect to the $\tilde{y}$-axis at $\phi = 0$; thus, the lower half is the mirror reflection of the upper half.  

In this region of the poloidal plane, once the conefield method for a given $t_f$ has identified a non-existence region $\tilde{\mathcal{S}}$, we found it sufficient to select four seed points in the complement of $\tilde{\mathcal{S}}$ to construct acceptable killends branches.
To be specific, the seed points are selected such that they lie on the boundary of $\tilde{\mathcal{S}}$, i.e., points $(\tilde{y}_i,\tilde{z}_j) \in \tilde{\mathcal{S}}^c$ such that at least one of $(\tilde{y}_i \pm \delta_y, \tilde{z}_j \pm \delta_z)$ belongs to $\tilde{\mathcal{S}}$, where $(\delta_y, \delta_z)$ are the generators of the grid.

These seed points are selected as the closest in the grid to the maximum amplitude of the magnetic islands, whose positions can be estimated from their corresponding resonances. In our case, the magnetic island associated with the $(2,1)$ resonance should have two lobules with maximum amplitude along the $y$-axis ($\vartheta = 0, \pi$) {in the chosen section}, while the magnetic island associated with the $(3,2)$ resonance is expected to have three lobules with maximum amplitude along the semi-lines with slopes $\vartheta = 0, 2\pi/3, -2\pi/3$.
 So the seed points are as follows: two points from the intersection of the inner component of $\partial \tilde{\mathcal{S}}$ with the $\tilde{y}$-axis, one point from the intersection of the exterior component of $\partial \tilde{\mathcal{S}}$ with the positive $\tilde{y}$-semiaxis, and the closest point in the exterior component of $\partial \tilde{\mathcal{S}}$ to the semi-line $\vartheta = 2\pi/3$. 

\begin{figure}[h!]
 \centering  
\includegraphics[height=7.7cm]{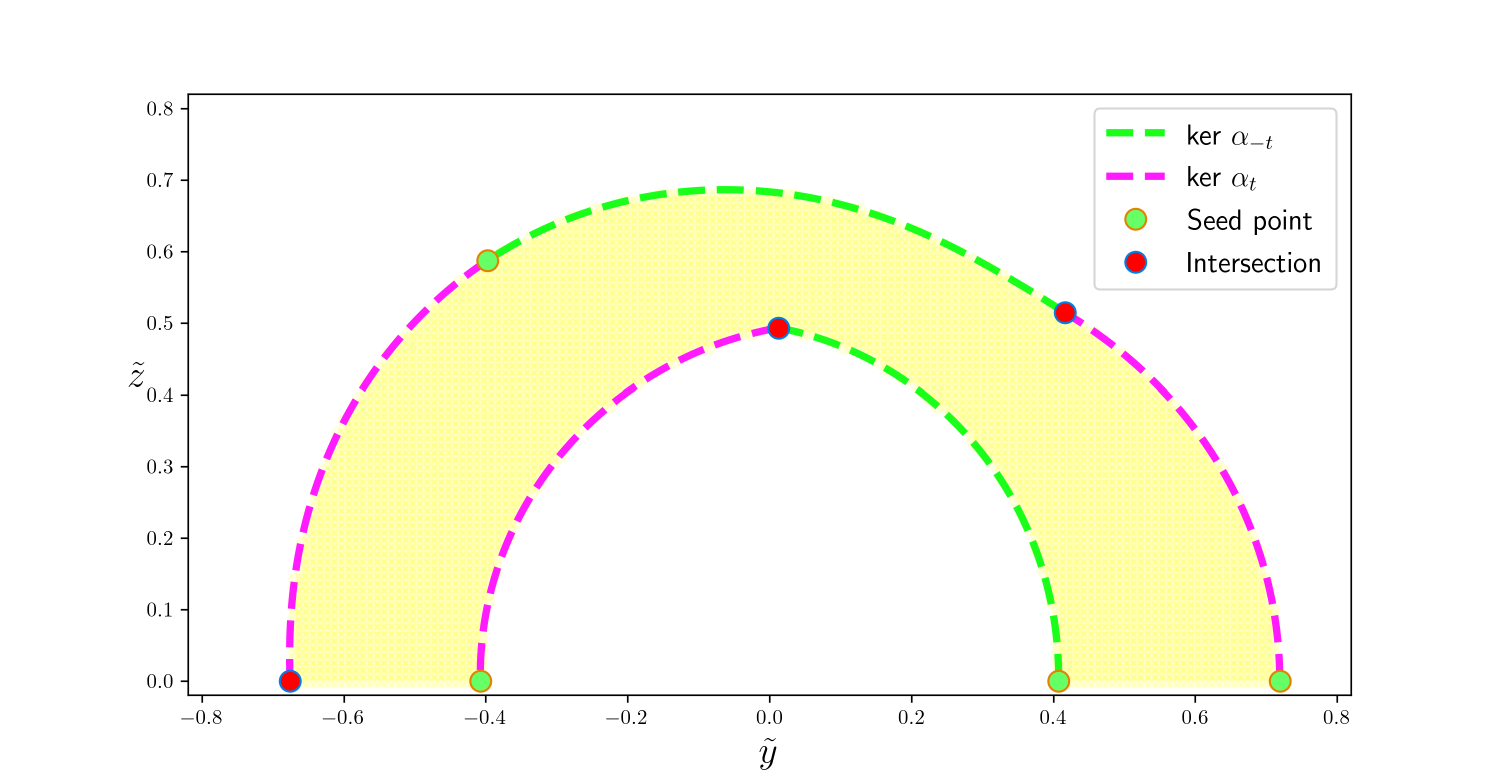}
\caption{Killends continuation branches and area regions for the upper plane of the example in symplectic coordinates $(\tilde{y},\tilde{z})$  for the perturbation parameter value $\varepsilon = 0.002$ and $t_f = 100$. The enclosed area is displayed in light yellow.}
   \label{Fig_area_enclosed}
\end{figure}

To measure the area enclosed by the killends branches,  
we used cubic spline interpolation on the points of the branches in $(\vartheta, \psi)$ coordinates to find the intersections between the branches and to identify the grid points contained within the enclosure formed by the five branches on the upper half-plane (see Figure \ref{Fig_area_enclosed}).
%////////////////////////////
%////////////////////////////

%////////////////////////////
\begin{figure}[h!]
 \centering  
\includegraphics[height=9.0cm]{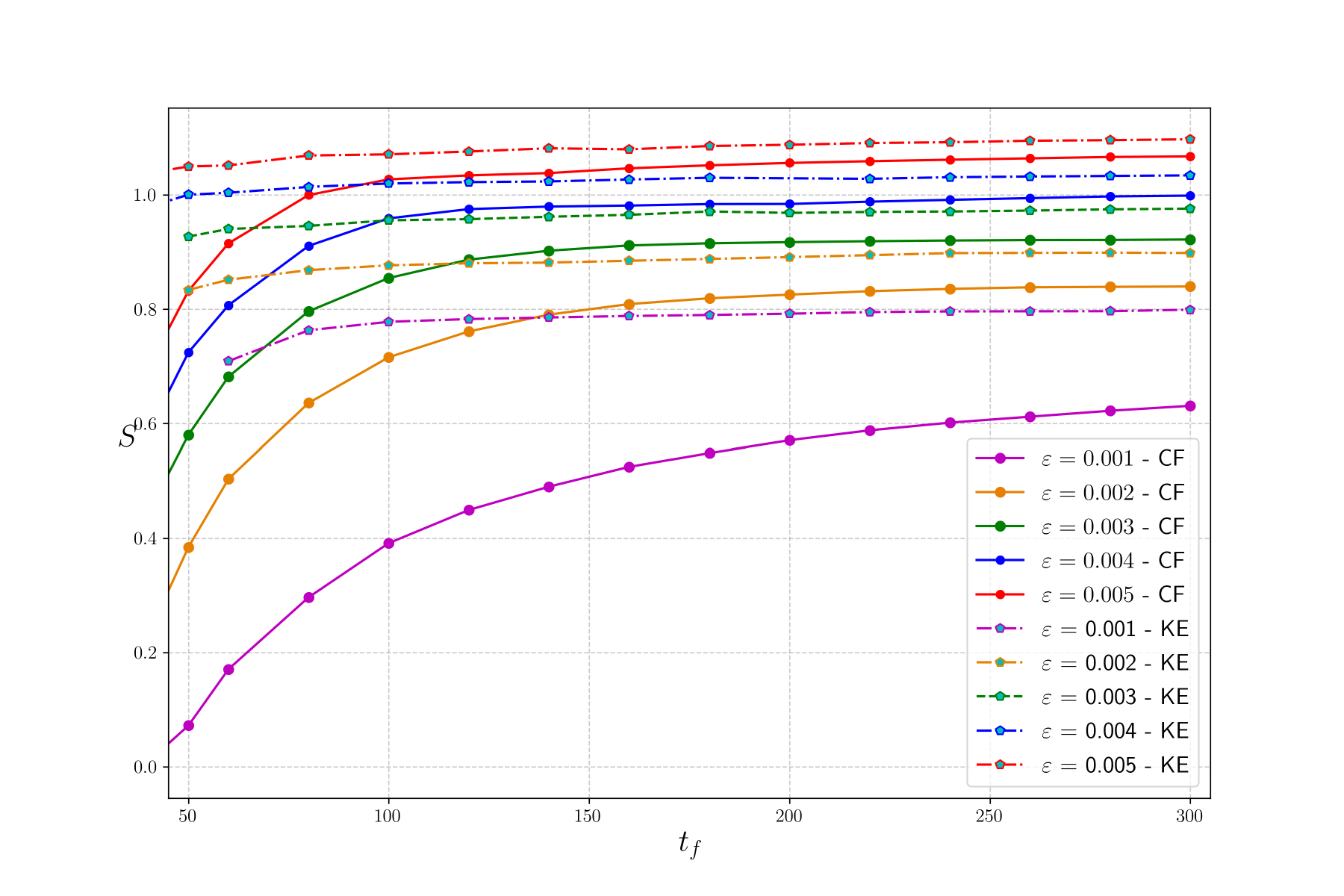}
\caption{Comparison of the killends enclosed area (KE) and non-existence area $S(t_f)$ detected by Converse KAM conefield method (CF) as a function of the timeout, for the upper plane of the example in symplectic coordinates for different values of $\varepsilon$.}
   \label{Fig_killends_enclosed2}
\end{figure}
%////////////////////////////

\begin{figure}[h!]
 \centering  
\includegraphics[height=7.7cm]{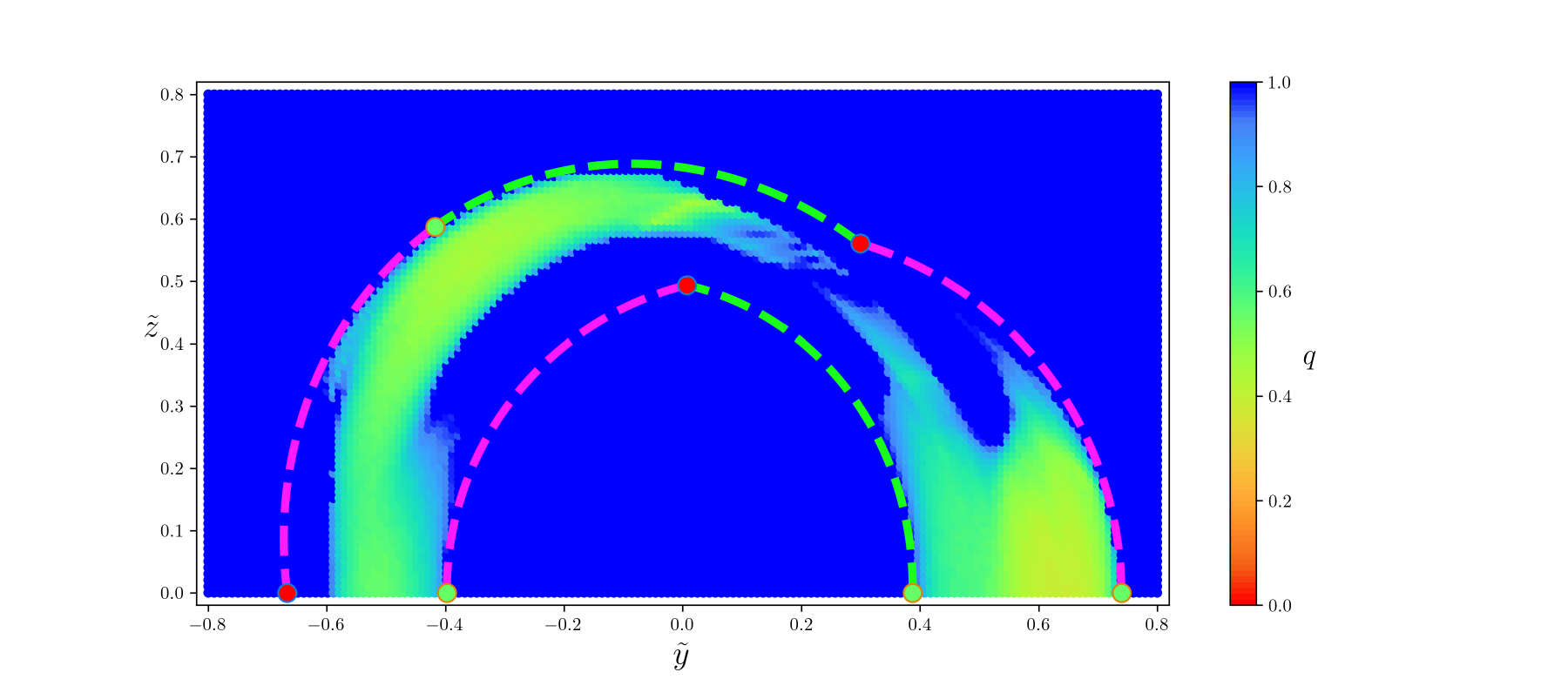}
\includegraphics[height=7.7cm]{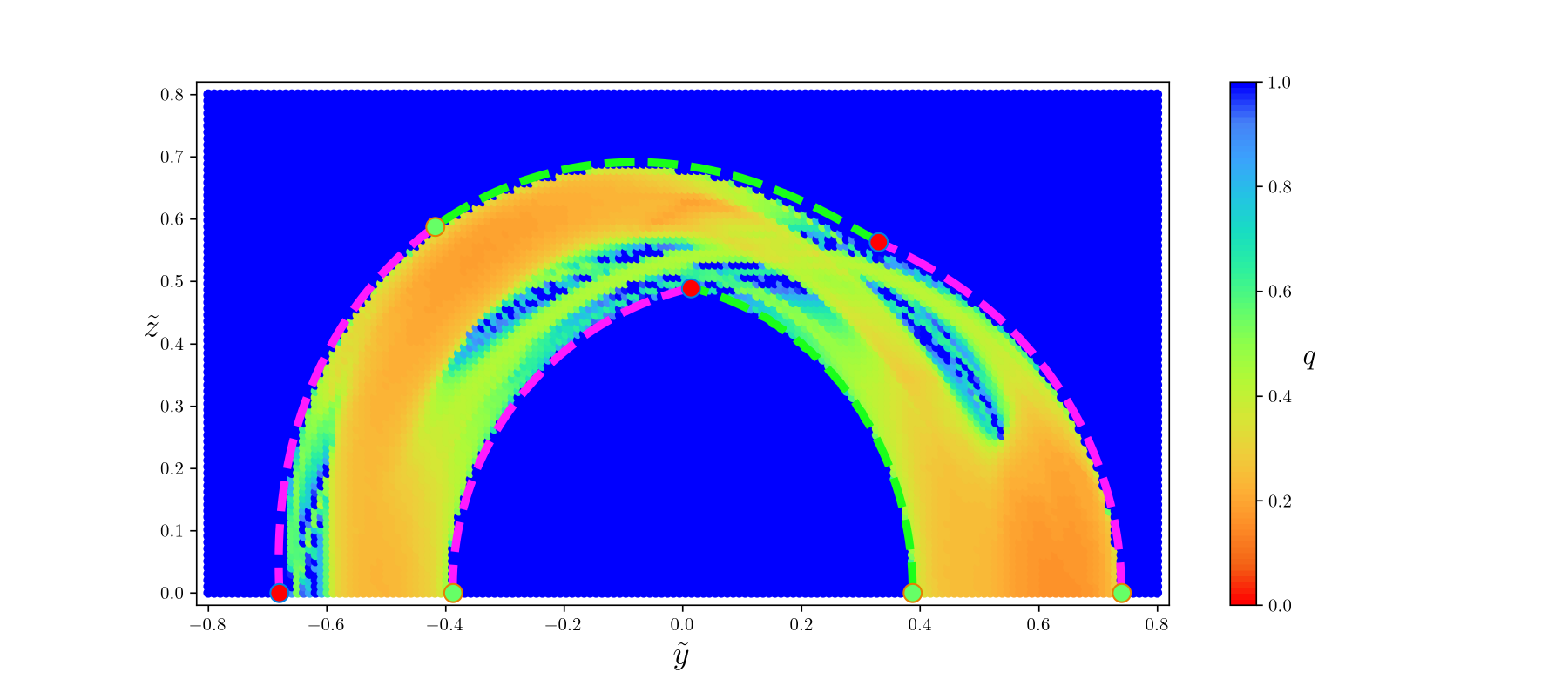}
\caption{Killends continuation branches superposed over the non-existence region detected by the conefield method over the upper plane of the example in symplectic coordinates $(\tilde{y},\tilde{z})$  for the perturbation parameter value $\varepsilon = 0.003$ and timeouts: $t_f = 50$ (top) and $t_f = 120$ (bottom).}
   \label{Fig_003_full}
\end{figure}

Figure \ref{Fig_killends_enclosed2} shows, for different values of $\eps$, the area of the non-existence region detected by the Converse KAM conefield method (solid lines), compared with the upper bounds found by the computation area enclosed by the killends branches (dashed lines).
The range of $t_f$ shown has as lower bound the minima for which the killends could be reliably computed from the conefield data, $t_f = 50$.
The figure suggests that the upper bound of the non-existence area found by the killends branches converges faster compared to the area eliminated by the conefield method and to a larger limiting value. The {improvement} can be attributed to structures within the non-existence region, where the dynamics are not favorable for the conefield method, such as cantori (see the hyperbolic invariant set that is apparent in the bottom pane of Figure \ref{Fig_003_full}), but which can be eliminated by killends.  
%////////////////////////////
%////////////////////////////

Indeed, in cases with positive shear, no fieldline of minimal action ($\int A\cdot dx$, considered as a functional of path in angular coordinates with $\psi$ determined implicitly) can be eliminated by the direct or conefield methods, and this includes not only those on invariant tori of the given class, but also the minimising hyperbolic periodic orbits and the fieldlines on minimising cantori{, but killends can eliminate minimising orbits if they do not lie on invariant tori.}
Thus, the killends extension to the conefield method makes an important contribution by eliminating more.  {Even though the area of the extra eliminated points might be negligible, the time taken to eliminate them is greatly reduced.}

\section{Discussion}
\label{sec:disc}

The magnetic fields we have studied here have a reflection symmetry (``stellarator symmetry'') \cite{dewar1998stellarator}. 
It could be used to speed up computations for initial conditions on the symmetry lines by a factor of two, as was done in \cite{mackay1985converse} and \cite{KMM23}.  We didn't implement it here, to avoid obscuring the general procedure.

The killends branches computation presented in this paper, used  some prior knowledge of the structures of the magnetic islands chosen.   For a more general case, in which the location or symmetry of the islands is not known beforehand, the implementation would require additional strategies to select suitable seed points. A challenge for the future is to automate the killends procedure.  This was achieved for area-preserving maps in \cite{mackay1985converse} by subdividing the domain into squares and computing upper and lower bounds on the slopes in each square; then killends was applied in an automatic way to eliminate whole squares where possible.  Something similar could be done here but it would be preferable not to take bounds on 
 a subdivision of the domain.

To simplify our presentation of the conefield method, we considered only differentiable tori of the given class.  But actually, the method applies equally well to all Lipschitz tori topologically transverse to the given direction field.  A torus is Lipschitz if it is locally the graph $z = f(x,y)$ of a Lipschitz function $f$, i.e.~for any point $(x_0,y_0,z_0)$ of the torus there is a neighbourhood in which the intersection with the torus is contained within $|z-z_0| \le L (|x-x_0|+|y-y_0|)$ for some $L$.  Then $\alpha^\pm$ constructed by the conefield method provide local Lipschitz bounds on the intersection of an invariant torus with a transverse section through the given point.
This is a particularly useful result in case of definite shear in a domain of the form of the product of a torus $\mathcal{T}$ with an interval, with direction field $\xi$ in the interval direction, because translating a theorem of Birkhoff (sec.44 of \cite{birkhoff1922surface}) 
to this context shows that every invariant torus that is homotopic to $\mathcal{T} \times {c}$, $c$ a constant, is transverse to $\xi$ and Lipschitz.  So the class of tori that the method treats can be considered to be all those homotopic to $\mathcal{T}\times {c}$.

As in \cite{KMM23}, the direction field used in the computations was radial, $\xi = \partial_\psi$.  This was chosen to eliminate points not on tori in the ``principal'' class that surround the magnetic axis, on the assumption that all such tori are transverse to the radial field.  As seen in the previous paragraph, this is justified in the case of definite shear.  Furthermore, changing the direction field a little does not change the definiteness of the shear, so does not change the class of tori eliminated.  But the assumption might not be justified in general.  For example, if the shear changes sign in the domain then there might be ``wiggly'' tori as found in 
\cite{howard1984stochasticity}. 

Moreover, it is possible to replace the radial direction field with a vector field that allows tori in  one or both of the magnetic islands in our examples too, as was done for an analogous problem in 
\cite{duignan2021nonexistence}. Our aim here was just to demonstrate use of the conefield method and killends extension in some simple cases; we leave more interesting applications for future work.

An important direction for future work is to apply the method to guiding-centre motion in a magnetic field.  This consists of a two-parameter family of 3D volume-preserving vector fields, with parameters the energy and the magnetic moment.  The direction field needs choosing to correspond to given class of guiding-centre motion. For volume-form on non-degenerate energy levels one can take $\frac{e\tilde{B}_\parallel}{v_\parallel} \Omega$, with notation as in 
\cite{burby2023isodrastic} (despite appearances this is non-singular).

\section{Conclusions}
\label{sec:conc}

In conclusion, we have reformulated the Converse KAM method that was applied to magnetic fields in \cite{KMM23}, in terms of construction of a conefield, and extended the non-existence region using the supplementary ``killends'' condition.  We find that using the killends extension we can eliminate more volume.  Furthermore, we find that we can eliminate more volume with less computation time. 

The conefield method is slightly more complicated to implement than the direct method of \cite{KMM23} but has the advantage of producing more information, namely the conefield, giving bounds on the slopes of all invariant tori of the given class and permitting the killends extension.  

The killends extension allows to eliminate many points more quickly and to eliminate some other points that would never be eliminated without it.  If implemented optimally, it is plausible that killends gets arbitrarily close to the true non-existence region as the timeout is increased (compare \cite{stark1988exhaustive}).  The results of our paper indicate that the non-existence region using killends saturates already for fairly short timeout in reasonable examples.

\section*{Acknowledgements}
We are grateful to Nikos Kallinikos for comments and to the reviewers for their careful reading, questions and suggestions.  This work was supported by a grant from the Simons Foundation (601970, RSM).

\subsection*{Code availability}
Code for the computations is available on
\url{https://github.com/dvmtz-1/cKAM-Conefield_implementation} 
\cite{cKAM2025}.

%\newpage
\bibliographystyle{unsrt}
\bibliography{ConvKAMbib}

\begin{thebibliography}{10}

\bibitem{dumas2014kam}
HS~Dumas.
\newblock {\em {The KAM Story: A Friendly Introduction to the Content, History,
  and Significance of Classical Kolmogorov-Arnold-Moser Theory}}.
\newblock World Scientific Publishing Company, 2014.

\bibitem{figueras2017rigorous}
J-Ll Figueras, A~Haro, and A~Luque.
\newblock {Rigorous computer-assisted application of KAM theory: A modern
  approach}.
\newblock {\em Foundations of Computational Mathematics}, 17(5):1123--1193,
  2017.

\bibitem{mackay1985converse}
RS~MacKay and IC~Percival.
\newblock {Converse KAM: Theory and practice}.
\newblock {\em Communications in Mathematical Physics}, 98(4):469--512, 1985.

\bibitem{mackay1989criterion}
RS~MacKay.
\newblock {A criterion for non-existence of invariant tori for Hamiltonian
  systems}.
\newblock {\em Physica D: Nonlinear Phenomena}, 36(1-2):64--82, 1989.

\bibitem{KMM23}
N~Kallinikos, RS~MacKay, and D~Martinez-del Rio.
\newblock Regions without flux surfaces of given class for magnetic fields in
  toroidal geometry.
\newblock {\em Plasma Physics and Controlled Fusion}, 65(9):095021, 2023.

\bibitem{mackay2018finding}
RS~MacKay.
\newblock {Finding the complement of the invariant manifolds transverse to a
  given foliation for a 3D flow}.
\newblock {\em Regular and Chaotic Dynamics}, 23(6):797--802, 2018.

\bibitem{kallinikos2014integrable}
N~Kallinikos, H~Isliker, L~Vlahos, and E~Meletlidou.
\newblock Integrable perturbed magnetic fields in toroidal geometry: An exact
  analytical flux surface label for large aspect ratio.
\newblock {\em Physics of Plasmas}, 21(6):064504, 2014.

\bibitem{mackay2020differential}
RS~MacKay.
\newblock Differential forms for plasma physics.
\newblock {\em Journal of Plasma Physics}, 86(1):925860101, 2020.

\bibitem{abdullaev2014}
S~Abdullaev.
\newblock {\em Magnetic Stochasticity in Magnetically Confined Fusion Plasmas}.
\newblock Springer, 2014.

\bibitem{stark1988exhaustive}
J~Stark.
\newblock An exhaustive criterion for the non-existence of invariant circles
  for area-preserving twist maps.
\newblock {\em Communications in mathematical physics}, 117:177--189, 1988.

\bibitem{dewar1998stellarator}
RL~Dewar and SR~Hudson.
\newblock Stellarator symmetry.
\newblock {\em Physica D: Nonlinear Phenomena}, 112(1-2):275--280, 1998.

\bibitem{birkhoff1922surface}
GD~Birkhoff.
\newblock Surface transformations and their dynamical applications.
\newblock {\em Acta Math.}, 43(1), 1920.

\bibitem{howard1984stochasticity}
JE~Howard and SM~Hohs.
\newblock Stochasticity and reconnection in hamiltonian systems.
\newblock {\em Physical Review A}, 29(1):418, 1984.

\bibitem{duignan2021nonexistence}
N~Duignan and JD~Meiss.
\newblock {Nonexistence of invariant tori transverse to foliations: An
  application of converse KAM theory}.
\newblock {\em Chaos}, 31(1):013124, 2021.

\bibitem{burby2023isodrastic}
JW~Burby, RS~MacKay, and S~Naik.
\newblock Isodrastic magnetic fields for suppressing transitions in
  guiding-centre motion.
\newblock {\em Nonlinearity}, 36(11):5884, 2023.

\bibitem{cKAM2025}
D~Martinez-del Rio.
\newblock c{KAM}-{C}onefield\_implementation.
\newblock \url{https://ggithub.com/dvmtz-1/cKAM-Conefield_implementation},
  \url{https://doi.org/10.5281/zenodo.14615555}, 2025.

\end{thebibliography}

\end{document}